\documentclass[twocolumn,amssymb,floatfix,deluxe,prb]{revtex4}

\usepackage{graphicx}
\usepackage{dcolumn}
\usepackage{bm}
\usepackage{epsfig}

\begin{document}

\title{\Huge\bf{Interpreting the X-ray Flash XRF 060218 and its associated supernova}}

\author{A. De R\'ujula}
\affiliation{Theory Unit, CERN,
1211 Geneva 23, Switzerland;
Physics Department, Boston University, USA}%



\maketitle

{\bf \noindent Forty years after their discovery, and in spite of a very
large body of observations, the operation of the `engine' responsible for 
long-duration Gamma-Ray
Bursts (GRBs) and X-ray flashes --as well as the mechanisms generating
their radiation-- are still the subject of debate and study. In this
respect a recent event\cite{218Disco}, XRF 060218, associated\cite{218SN} with 
SN 2006aj, is particularly significant. It has been argued that, for the
first time, the break-out of the shock involved in the supernova explosion
has been observed, thanks to the detection of a thermal component in
the event's radiation\cite{Campana}; that this XRF was not a GRB seen
`off-axis', but a member of a new class of energetically
feeble GRBs\cite{Amati218}; and that its `continued engine activity' may have been driven 
by a remnant highly-magnetized neutron star, a magnetar\cite{PianMagnetar}. 
I argue, on grounds based on observations and on limpid verified hypothesis, 
that there is a common, simpler alternative to these views, with 
no thermal component, no new feeble GRBs, and no steady engine activity.
}

Many astrophysical systems, such as quasars and micro-quasars, emit relativistic
jets. The `engines' generating `long' GRBs are the supernovae (SNe) `associated'
with them. The radiation we perceive as a GRB cannot be isotropic, if only because
the available energy in a SN event is insufficient. Thus, GRBs must be beamed,
and it is natural to assume that the source of their radiation is a jet, emitted by
their engine SN. A highly relativistic beam suggests itself for, as a simple consequence 
of special relativity, its radiation is highly forward-collimated and boosted in energy. 
The beam consists of a succession of ejecta, for a GRB's radiation typically consists
of an aperiodic succession of `pulses'. Before the interactions with the ambient
matter significantly affect it, the ejected matter responsible for a pulse must trace
a cone in its voyage, as it freely expands (in its rest system) and moves inertially
in space. Up to this point, practically all interpretations of GRBs are 
concordant\cite{Meszaros,ADR}. 

Let $\theta_j$ be the (half) opening angle of the cone we cited, and let $\alpha$ be
the angle, relative to the cone's  axis, of a given radiation-emitting point, which 
moves with the common bulk Lorentz factor, $\gamma$, of 
the emitting region ($\gamma\!=\!1/\sqrt{1-\beta^2}$; $\beta\!\equiv\! v/c$). 
The probability of an observer to be located in the direction $\alpha$, with a
precision $d\alpha$, is $dP\! \propto\! \alpha\, d\alpha$. On axis, $dP\!=\!0$.
Since $\alpha\le\theta_j$, $dP$ (whose integral is quadratic)
is maximal at the cone's rim, $\alpha\!=\!\theta_j$. This is where located GRBs 
must be concentrated.

The probability to detect events does not vanish abruptly at $\alpha\! >\! \theta_j$,
since a point's radiation is forward collimated, but not infinitely so. The degree
of collimation and the energy boost of the radiation from a given point
are dictated by the Doppler factor 
$\delta(\theta_p)\!=\!1/[\gamma(1-\beta\cos\theta_p)]$, with $\theta_p$ the angle
between the point's direction of motion and the observer's direction. 
For $\gamma\!\gg\! 1$ and $\theta_p\!\ll\! 1$, to an excellent approximation,
$\delta(\theta_p)\!=\!2\,\gamma/[1+(\gamma\,\theta_p)^2]$. To be fully precise,
I ought to average over the distribution of emitting points\cite{DDDcorrels}, 
but some trivial limiting cases will suffice here.

For observer's directions, $\theta_{\! ob}$, beyond the cone's rim ($\theta_{\! ob}\!>\!\theta_j$)
the point-averaged $\langle\delta(\theta_p)\rangle$ rapidly tends to $\delta(\theta_{\! ob})$, the
limit for a point-like source. The properties of GRBs seen at such angles,
for which $\delta$ is a rapidly diminishing function of $\theta_{\! ob}$,
are easy to predict. Relative to the average GRB, their `isotropic equivalent' 
energy, $E_{\rm iso}\!\propto\!\delta^3$, as well as the `peak energy' 
of the photons in their pulses, 
$E_p\!\propto\!\delta/(1+z)$, must be small, declining in a correlated way
(to be precise, I introduced the effect of the cosmological redshift, $z$).
The time-widths of XRF peaks, $\Delta\!\propto\! (1+z)/\delta$ must be
relatively large. The peak-to-peak intervals, still relative to GRBs, are not
affected; the engine determines them, and it is not moving. Since what I have described 
are the observed properties of XRFs, it is natural to propose that they are GRBs seen 
off-axis\cite{DD2004,XRFs}, even if the argument is less
compelling in the $\theta_j\!\gg\!1/\gamma$ limit. The
traditional distinction between XRFs and GRBs is that the former have
$E_p\!<\!50$ keV.

I shall assume that XRF 060218, which has all the XRF properties
I have described, is also an ordinary GRB seen off-axis. 
As we proceed, this view will gain support. 
The opposite view\cite{Amati218} will be discussed anon. 

To proceed, we need some observational input. The spectrum of the `prompt'
radiation of GRBs and XRFs is well described by the `Band' function\cite{Preece}. 
For the relatively low X-ray energies of interest here, the first addend of this
function, a `cutoff power-law' (CPL), suffices:
\begin{equation}
E \,dN_\gamma/ dE\vert_{\rm CPL} = a\,E^\epsilon\,{\rm Exp} ( - E / E_p ),
\label{Band} 
\end{equation}
where $E$ and $N_\gamma$ are photon energies and numbers,  $a$
is  a normalization and $\epsilon\!\sim\!0$ is a good approximation
case by case, a very good approximation on average\cite{Preece}.
The peak energy of individual photons
decreases during a pulse, rapidly tending to $E_p(t)\!\approx\!b / t^2$,
with $b$ a case-by-case parameter.
This correlation between times and energies is better known and
studied in its complementary forms: the increasing  `lag-times' of the peaks of
GRB pulses in decreasing energy intervals\cite{WuFen},
and the relation between pulse widths and energy intervals\cite{Norris}, 
roughly $\Delta\!\propto\!t^{-1/2}$. Taking into account the time dependence
of $E_p$, and choosing $\epsilon\!=\!0$,
we may expect the spectrum of a particular GRB pulse to be roughly  described by:
\begin{equation}
E\, dN_\gamma/ (dE dt) = a\,{\rm Exp} (-E\, t^2 / b ).
\label{spectrum}
\end{equation}

\begin{figure}
\centering
\epsfig{file=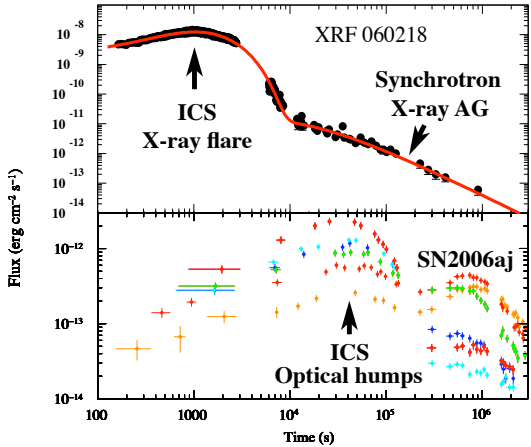,width=10cm}
\vskip -1.3cm 
\caption{{\bf (a)} {\it Upper panel}: the  (0.3--10 keV) X-ray 
flux\cite{Campana,Soderberg,deLuca} of XRF 060218. 
{\bf (b)} {\it Lower panel}: the UVOT flux (from Ref.~3),
for different UVOT filters (from bottom to top between $10^4-10^5$ s): 
V, B, U, UVW1, UVM1,
and UVW2. 
}
\vspace{-.5cm}
\label{f1}
\end{figure}

\begin{figure}[]
\centering
\vbox{
\epsfig{file=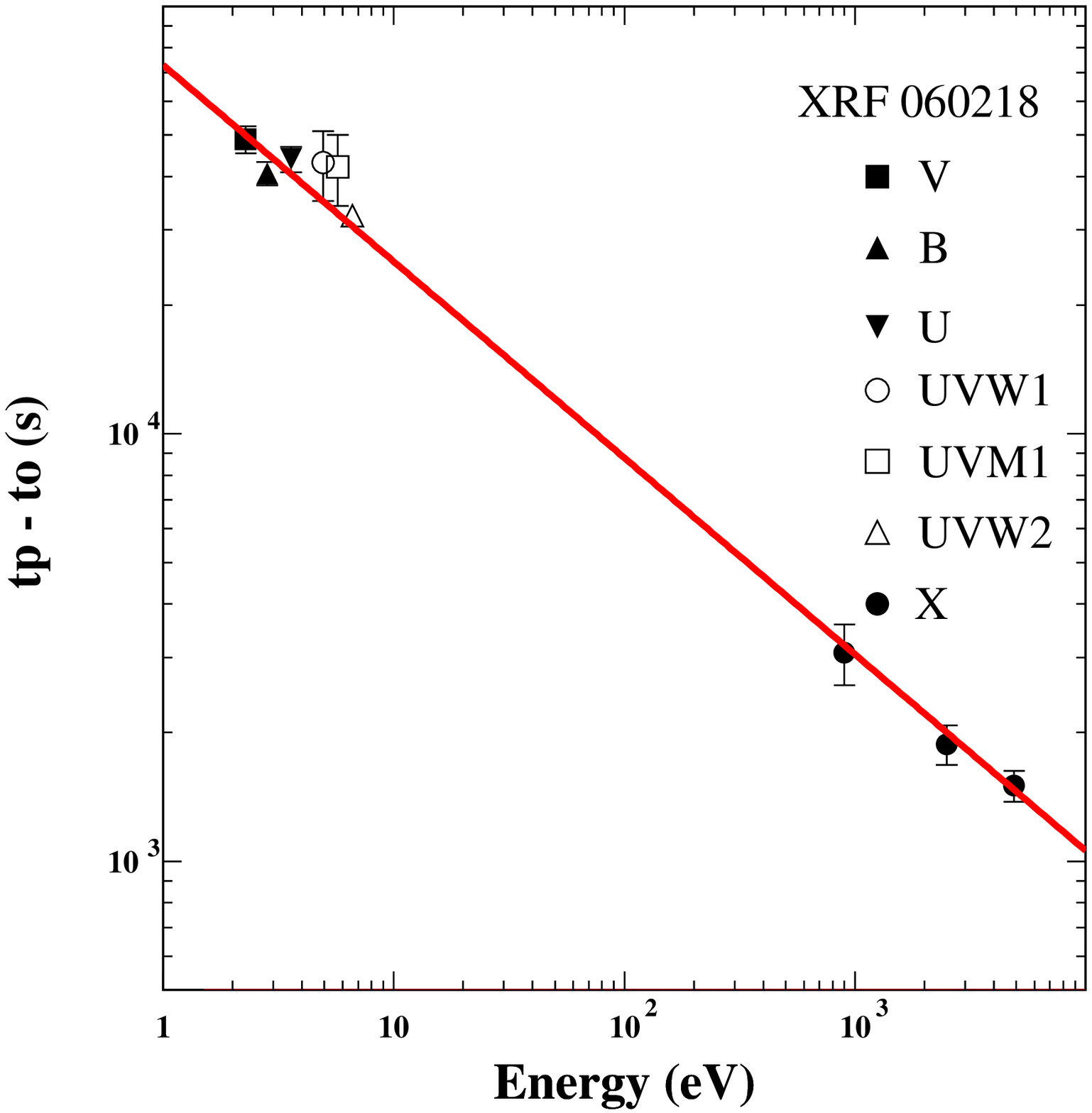,width=6.5cm}
\epsfig{file=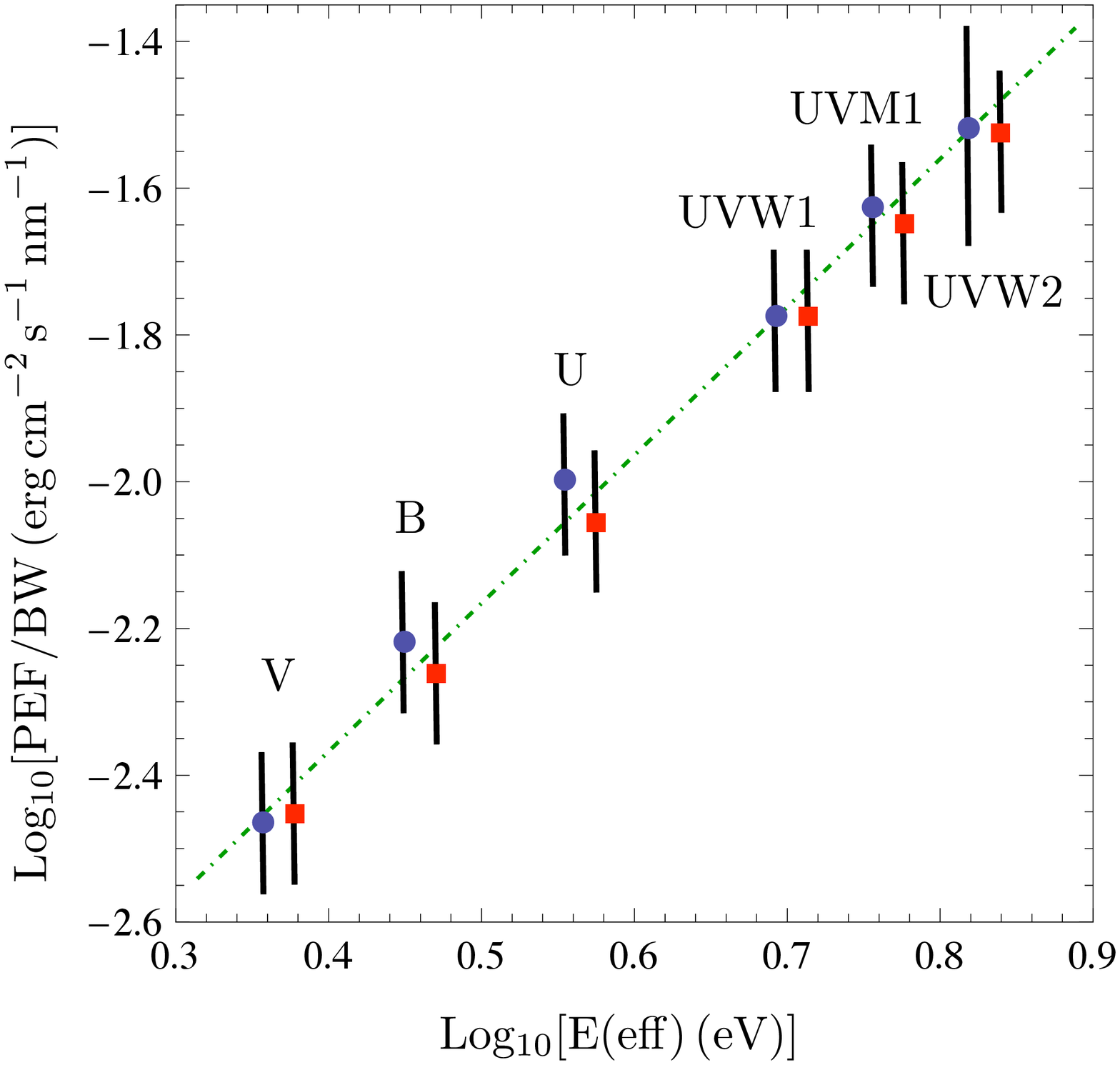,width=5.9cm}
}
\vspace{-3cm}
\caption{{\bf (a)} Comparisons of Eq.~(\ref{spectrum}), with data
corrected for extinction\cite{DDDXrays}.
{\it Upper panel}: The `$E\,t^2$ law', with $t_p\!-\!t_0$ the
peak position at various energies and $t_0$ the common peak start-time at all $E$.
The line is the expected $t_p\!-\!t_0\!\propto\!1/\sqrt{E}$.
{\bf (b)} {\it Lower panel}: The spectral dependence.
Data and predictions are slightly shifted, to avoid superposition. At UVOT
energies, the exponential in Eq.~(\ref{spectrum}) is $\approx\!1$ and 
$d\,{\rm PEF}/d\lambda\!\propto\!E^2$, the dashed line, whose normalization is
fixed by the X-ray data. 
}
\vspace{-.6cm}
\label{f2}
\end{figure}

Having posited that XRF 060218 is but a GRB seen off axis, it behooves me to
test whether it satisfies Eq.~(\ref{spectrum}). The X-ray\cite{Campana,Soderberg,deLuca} 
and optical\cite{Campana} 
observations of this single-peak event are shown in Fig.~\ref{f1}. The top part
is the X-ray flux in the cumulative (0.3--10) keV interval, fit in a particular model\cite{DDDXrays},
in which Eq.~(\ref{spectrum}) is the approximate expectation\cite{DD2004}. In 
Fig.~\ref{f2}a,
I test the `$E\, t^2$ law' implied by Eq.~(\ref{spectrum}), by plotting (versus energy)
the peak times of the optical and UV `humps' of Fig.~\ref{f1}b, and of their
sister X-ray peak, measured in three energy sub-intervals. In Fig.~\ref{f2}b,
I show the test of the spectral behaviour of  Eq.~(\ref{spectrum}).
For lack of space, the three X-ray intervals, whose relative spectrum satisfies
Eq.~(\ref{spectrum}), are not shown. Using them as normalization, one can
predict the `peak energy fluxes' (per unit wave-lengh) in the optical
and UV channels, and compare them, as in Fig.~\ref{f2}b, with the observations.
The results are fairly satisfactory.

In the standard interpretation\cite{Campana} of this XRF, a thermal spectrum 
$E\, dN_\gamma/ dE\!\propto\! R^2(t)\,E^3\, {\rm Exp} [- E / T(t) ]$
is added to the CPL spectrum of Eq.~(\ref{Band}), with a varying
radius and temperature, $R(t)$ and $T(t)$, fit, bin by bin, to the
observations. The CPL spectrum masks the tell-tale $E^3$
thermal dependence. The time dependence of the extracted $T(t)$ is
roughly $\propto\!1-t/\sqrt{t^2+\Delta^2}$, which is the expected
dependence of $E_p(t)$, to a better approximation than 
$E_p(t)\!\propto\! t^{-2}$, in the model I have quoted\cite{DD2004}.

Had Campana et al.~\cite{Campana} used a time-dependent
$E_p(t)$, they would have obtained the results in Fig.~\ref{f2},
and, I presume, concluded that a thermal component is unnecessary. 
These authors cite Amati et al.\cite{Amati218} for the assertion
that XRF 060218 is not a GRB seen off axis, but a representative
of a new class of sub-energetic GRBs. Presumably that is why what
is known about $E_p(t)$ in `normal' GRBs was not judged to be
relevant. 

Amati et al.\cite{Amati218} base their cited conclusion on the following
fact. If one makes a scatter log-log plot of the $E_p$ values versus 
total fluence\cite{Lloyd} of a collection of GRBs, one finds
that they fall, with not much scatter, close to a straight line. 
A similar result is obtained\cite{Amati2} for a log-log $[(1+z)\,E_p,E_{\rm iso}]$ 
plot of GRBs of known $z$, the `Amati correlation'. These results are
`phenomenological', the lines' slopes and normalizations are fit
to the results. 

To understand the origin of the correlations between `prompt'
observables, such as $E_p$ and $E_{\rm iso}$, one must specify the
radiation mechanism. Let us posit that it is inverse Compton
scattering (ICS) of `ambient' photons by the electrons in a GRB's jet\cite{ShavivDar},
as verified in Ref. 9. In that case,  the $\delta$-, {\it and}
$\gamma$-dependences of $E_p$ and $E_{\rm iso}$ can be specified.
For a point-like source, they are\cite{DD2004} $(1+z) \,E_p\! \propto\! \gamma\, \delta$
and $E_{\rm iso}\! \propto\! \delta^3$. Since $\delta$ decreases, at fixed $\gamma$,
by orders of magnitude as $\theta_{\! ob}$ increases, and $\gamma$ cannot be much
smaller than `typical' (for otherwise we would not be observing a GRB
or an XRF), we conclude that the case-by-case variability is dominated
by $\delta$. Thus, we expect $(1+z) \,E_p\! \propto\!(E_{\rm iso})^{1/3}$.
Since we used the point-like limit, valid for large $\theta_{\! ob}$, this result is 
for XRFs.

\begin{figure}
\includegraphics[width=8cm]{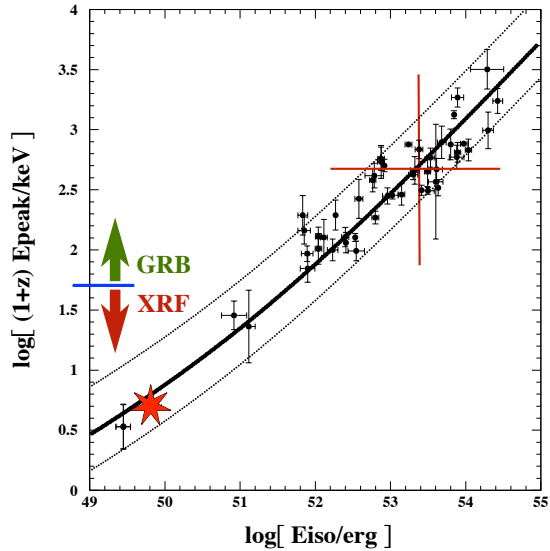}
\caption{The $[(1+z)\,E_p,E_{\rm iso}]$ correlation\cite{DDDcorrels}.
XRF 060218 is the star. The cross marks the average expectation for GRBs\cite{DD2004}.
}
\vspace{-.5cm}
\label{f3}
\end{figure}

In the opposite extreme at which a GRB is seen at an angle $\theta_{\! ob}\!\ll\!\theta_j$,
the integration over the source's radiating points masks the dependence
on $\theta_{\! ob}$, so that the `effective' averaged Doppler factor\cite{DDDcorrels}
becomes $\delta\!\propto\!\gamma$.
Thus, for hard GRBs,
$(1+z) \,E_p\! \propto\!\gamma^2$ and  $E_{\rm iso}\!\propto\! \gamma^3$,
implying that $(1+z) \,E_p\! \propto\!(E_{\rm iso})^{2/3}$. Since the observer's
angle varies continuously, we expect\cite{DDDcorrels} the XRFs and GRBs to lie, in
the $[(1+z)\,E_p,E_{\rm iso}]$ log-log plane, close to a line whose slope
varies smoothly from $1/3$ to $\sim\!2/3$. This is tested in Fig.~\ref{f3}. The crossing lines
are the predicted average values for GRBs, for the advocated
origin of the target light that is Compton 
up-scattered\cite{DD2004}. XRF 060218 is the star in Fig.~\ref{f3}.
Incidentally, the above reasoning also explains the observed correlations with
(or between) other prompt observables: peak luminosity, lag-time, rise-time
and variability\cite{DDDcorrels}.
GRB 980425 is an out-lier from correlations involving $E_p$, its `second $E_p$', 
due to ICS by
electrons scattered by the jet\cite{DD425}, is what was observed.

We have been led to conclude that XRF 060218 {\it is} a  normal
GRB seen off-axis, among other reasons, {\it because} it satisfies the correlation 
we discussed. Amati et al.\cite{Amati218} conclude that XRF 060218 {\it is not} a  
normal GRB and {\it is not} seen off-axis, {\it because} it lies in a similar correlation.
This logical inconsistency, I believe, has a purely logical origin.
Their argument\cite{Amati218} is the following:
GRBs are seen close-to-axis, GRBs satisfy the Amati correlation, XRF 060218 
does it as well, so this object is seen on axis and, since it has very low $E_p$
and $E_{\rm iso}$, it belongs to a new class. My critique of this logic
is that, as I argued at the beginning, GRBs are not seen on axis, but on-edge.
Since the opposite premise is tacit in Amati et al.\cite{Amati218}, one cannot be
sure this interpretation of their logic is correct.

In the X-ray flux shown in Fig.~\ref{f1}a, there is a `plateau' tail labeled 
`Synchrotron Afterglow', fit, as the rest of the curve, to a given model\cite{DDDXrays}.
The whole curve has a `canonical' shape, sketchily seen long ago in GRB 980425, 
and nine others GRBs\cite{DDD2002}. The X-ray and radio afterglows (AGs) of GRB 980425
were attributed to a non-relativistic shock\cite{Waxman425,WaxmanLoeb}. 
Alternatively, since they coincide with the expectation for an off-axis relativistic
jet of an otherwise normal GRB, they were
attributed to the jet\cite{DDD2002,DDD2003}. A large fraction of the X-ray AGs recently observed by Swift
and other satellites are `canonical'. Their plateau phases are, as in the case\cite{PianMagnetar} of
XRF 060218, attributed to a continued engine activity\cite{Nousek},
perhaps driven by a magnetar remnant\cite{Nakamura,PianMagnetar}.
In the alternative view, once again, the plateau is due to the 
moving jet\cite{DDDCanonical}. 
The difference between these views may be important, not only
because of the different underlying GRB theories, but because of its implications
on the associated SNe. Suffice it to imagine\cite{DDD2002,DDD2003}
that the X-ray and radio emissions attributed
to a SN were actually emitted by a relativistic jet having long departed from
the SN location.

The algebra required to discuss GRB afterglows is a bit longer than the
one I have been using. For this reason, I shall next state
some results with citations, but no proof.
The `alternative view' I  
repeatedly quoted\cite{DDD2002,DD2004} does not require, so far,
any conceptual changes or
additions, as it confronts the data. The steady `continued engine activity',
for instance, is instead the inevitable inertia of the relativistic jet, its ending
`break' occurring when the interstellar medium begins to significantly decelerate 
the jetted material\cite{DDD2002}. The rapidly-varying spectra of X-ray light curves, 
as well as the plethora of different Swift light-curve shapes, 
with and without `breaks', were not
unexpected\cite{DDDfast,DDDbreaks}.  In the
realm of `fireball models', very many novel phenomena --other than the
shock break-out, the steady engine activity, and the subenergetic GRBs that I have
discussed here-- have been invoked to understand GRBs and XRFs. 
To each of these phenomena, the `other view' I have often 
cited --{\it the  Cannonball Model}-- offers a simpler alternative. And it is always 
the same.

Ab initio, the `firecone' and cannonball models differed in almost 
all respects: supernovae as engines (debated {\it vs} assumed),
the substance of the ejecta ($e^+\,e^-$ pairs {\it vs} ordinary matter),
the role of shocks (decisive {\it vs} absent), the mechanism of $\gamma$-ray 
generation (synchrotron radiation {\it vs} ICS), to name a few\cite{Meszaros,ADR}. One difference 
still remaining underlies the conflicting views I have discussed.
It is the jet's initial opening angle, $\theta_j$: a few degrees (as in radio images of quasars),
{\it vs} $\sim\!1$ milli-radian. Moreover, a $\theta_j(t)$ that increases  with time, {\it vs} one that 
eventually decreases. The last and apparently\cite{DDD2002,DD2004}
surprising feature --a jet that does not expand--
is observed in sharper (X-ray) images of quasars, e.g.~Pictor A (Ref. 30).

The fraction of SNe that are associated to GRBs and XRFs is $\propto\!\theta_j^{-2}$,
differing by $\sim\!300$ for a one degree {\it vs} a one milli-radian angle. For 
$\theta_j\!\sim\! 1$ mrad the fraction of SNe associated with GRBs is, within
rather large `cosmological' uncertainties, close to {\it all} Type Ic SNe\cite{DDD2002,DD2004},
very far from the standard view\cite{Soderberg}. This is one
more reason why the opening angle is so decisive.
The value of $\theta_j$ adopted in the cannonball model corresponds
to $\theta_j\!\sim\!1/\gamma$, i.e.~to the cone swept by a object which, in its rest system,
expands at the speed of light, or of relativistic sound ($c/\sqrt{3}$), and is
moving with the typical initial $\gamma\!\sim\!10^3$ required by the data\cite{DDD2002,DD2004}.
For the same reason, the quest for simplicity, the original fireball models had spherical 
symmetry\cite{Meszaros}. The correct answer must lie in between these two extremes, 
surely closer to the former than to the latter.

\noindent
{\bf Aknowledgements.} I am indebted to my colleagues Shlomo Dado and Arnon Dar for
a long collaboration.

\end{document}